\documentclass[lineno]{jfm_arxiv}
\newcommand{\RomanNumeralCaps}[1]
\usepackage{graphicx,graphics}
\usepackage{newtxtext}
\usepackage{newtxmath}
\usepackage{natbib}
\usepackage{hyperref}

\newcommand{\bminil}[1]{\begin{minipage}[l]{#1 \textwidth}}
\newcommand{\bminir}[1]{\begin{minipage}[r]{#1 \textwidth}}
\newcommand{\bminic}[1]{\begin{minipage}[c]{#1 \textwidth}}
\newcommand{\emini}{\end{minipage}}
\newcommand{\EQL}{\begin{equation}\label}
\newcommand{\EQ}{\begin{equation}}
\newcommand{\EN}{\end{equation}}
\newcommand{\BFG}{\begin{figure}}
\newcommand{\EFG}{\end{figure}}
\newcommand{\ITM}{\begin{itemize}}
\newcommand{\ITN}{\end{itemize}}
\newcommand{\EEM}{\begin{enumerate}}
\newcommand{\EEN}{\end{enumerate}}
\newcommand{\BEA}{\[\begin{array}}
\newcommand{\EEA}{\end{array}\]}
\newcommand{\EQAL}{\begin{eqnarray}\label}
\newcommand{\EQA}{\begin{eqnarray}}
\newcommand{\ENA}{\end{eqnarray}}

\newcommand{\btriangle}{\mbox{\boldmath$\triangle$}}

\newcommand{\bS}{\mbox{\boldmath$S$}}

\newcommand{\cL}{{\cal L}}

\newcommand{\bk}{\mbox{\boldmath$k$}}

\newcommand{\br}{\mbox{\boldmath$r$}}
\newcommand{\bs}{\mbox{\boldmath$s$}}

\newcommand{\bu}{\mbox{\boldmath$u$}}

\newcommand{\bomega}{\mbox{\boldmath$\omega$}}

\newcommand{\bxi}{\mbox{\boldmath$\xi$}}

\newcommand{\ppto}[1]{\frac{\partial #1}{\partial t}}

\newcommand{\half}{\mbox{$\frac{1}{2}$}}

\renewcommand{\theenumi}{\alph{enumi}}

\newcommand{\cH}{{\cal H}}
\newcommand{\cI}{{\cal I}}
\newcommand{\etall}{{\it et al.}}
\newcommand{\biband}{\&~}
\newcommand{\authone}[2]{#2,~#1}
\newcommand{\authtwo}[4]{#2,~#1,~\&~#4,~#3}
\newcommand{\auththr}[6]{#2,~#1,~#4,~#3,~\&~#6,~#5}
\newcommand{\authfour}[8]{#2,~#1,~#4,~#3,~#6,~#5,~\&~#8,~#7} 
\newcommand{\authmanytwo}[4]{#2,~#1,~#4,~#3,} 
\newcommand{\authmanythr}[6]{#2,~#1,~#4,~#3,~#6,~#5,}

\newcommand{\yjour}[6]{ #1~ #6. {\em #2} {\bf #3}, #4#5.}

\title{\large Generating Kolmogorov spectra and structure functions with a
Rosenhead profile trefoil vortex knot}
\par\vspace{-5mm}
\author{Robert M. Kerr}

\affiliation{University of Warwick, Coventry CV4 7AL, UK}

\pubyear{2026}
\volume{}
\pagerange{}

\begin{document}
\maketitle
\par\vspace{-5mm}
\begin{abstract} 
Motivated by recent experimental observations by \cite{MatsuetalPNAS2026} of 
colliding several vortex rings within a water tank, this paper extends a compact 
trefoil vortex simulation to later times and generates a persistent $k^{-5/3}$ energy spectrum that obeys Kolmogorov scaling, together with other statistics characteristic of fully developed turbulence, including structure function scaling. 
This temporal extension uses a larger $(4\pi)^3$ periodic domain, with the improved 
compactness being achieved by the use of an algebraic Rosenhead vorticity profile,
from which viscosity($\nu$)-independent finite energy dissipation is generated
and pre-reconnection convergence of $\nu^{1/4}$-dependent higher-order vorticity moments. These properties suggest, perhaps for the first time, how decaying Navier-Stokes simulations can generate fully-developed turbulent flow from an unforced initial state that is far from boundaries, thereby suggesting how further simulations at higher values of $Re$ could be harnessed for experimental comparisons. Moreover, the exponents for the energy decay and the integral scale growth, $E(t)\sim(t+t_0)^{-10/7}$ and $\ell(t)\sim(t+t_0)^{2/7}$, follow those based upon preservation of the Loitsyanskii integral. On a related question, an integral scale growth rate $\ell$ that is close to
the experiment rate of $\ell_X\sim(t+t_0)^{0.16\approx 1/7}$ is only obtained if 
the Fourier modes at the scale of the periodic domain are not considered. 

\end{abstract}
\section{Background}\label{sec:back}
One of the greatest challenges in improving our understanding of three-dimensional
incompressible turbulence is producing, without external forcing, a turbulent flow that 
is far from boundaries, has a cascade of energy to small scales, 
finite dissipation rates and
a $E(k,t)\sim k^{-5/3}$ wavenumber spectrum.  Unlike turbulence associated with boundary layers \citep{SmitsARFM2011}, or with thermal convection 
\citep{BodenschatzARFM2000}, a clear experimental equivalent has not been found, with only measurements from wind tunnels and wakes available for comparative purposes. Moreover, numerical comparisons have been limited to isotropic calculations in periodic domains
across periodic boundaries that require the use of artificial forcing to sustain 
any turbulence.
\par\smallskip
In a recent significant experiment, turbulence far from boundaries has been generated by the simultaneous firing of multiple vortex rings into a water tank \citep{MatsuetalPNAS2026}, from which $E(k,t)\sim k^{-5/3}$ spectra have been observed over two  wavenumber decades that persist for several characterisic time-scales, or eddy turnover times. 
\par
\begin{figure} \hspace{-7mm}
\includegraphics[scale=.160]{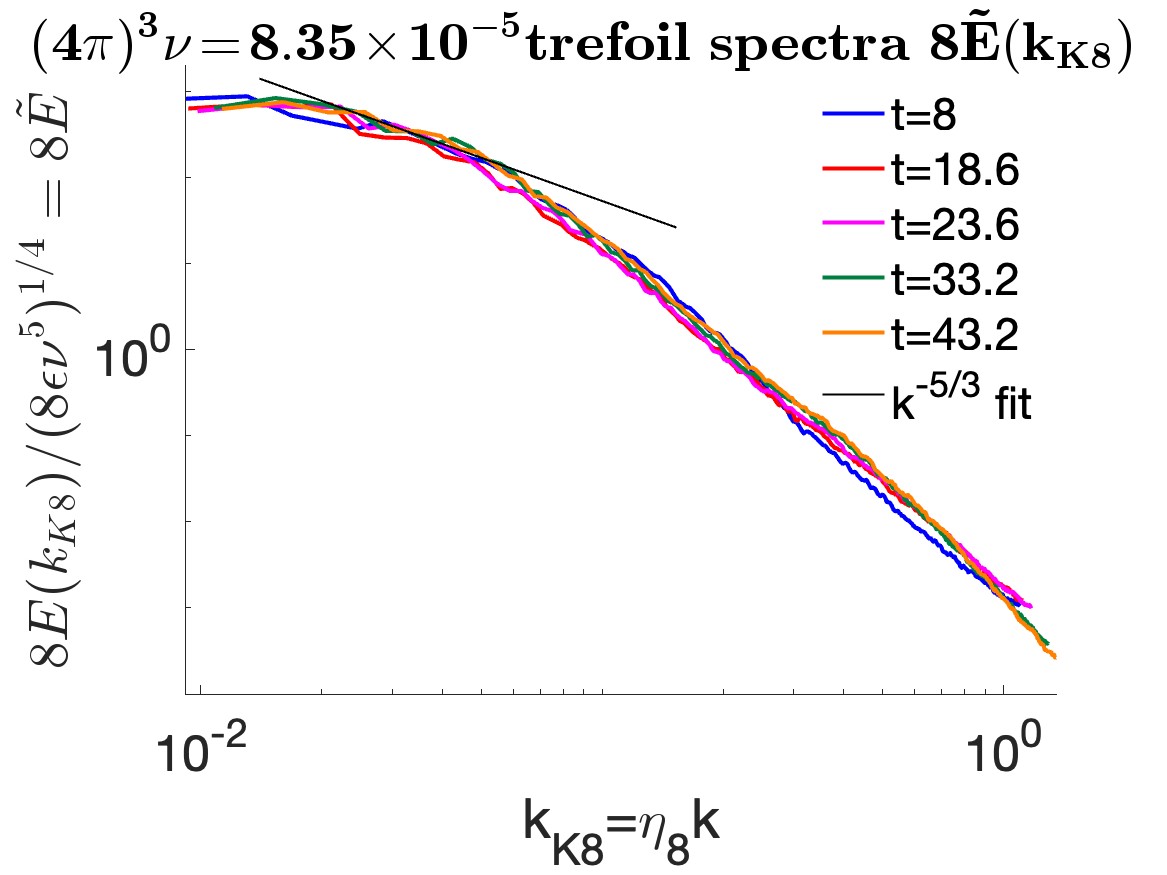}
\includegraphics[scale=.160,clip=true,trim=0 0 0 0]{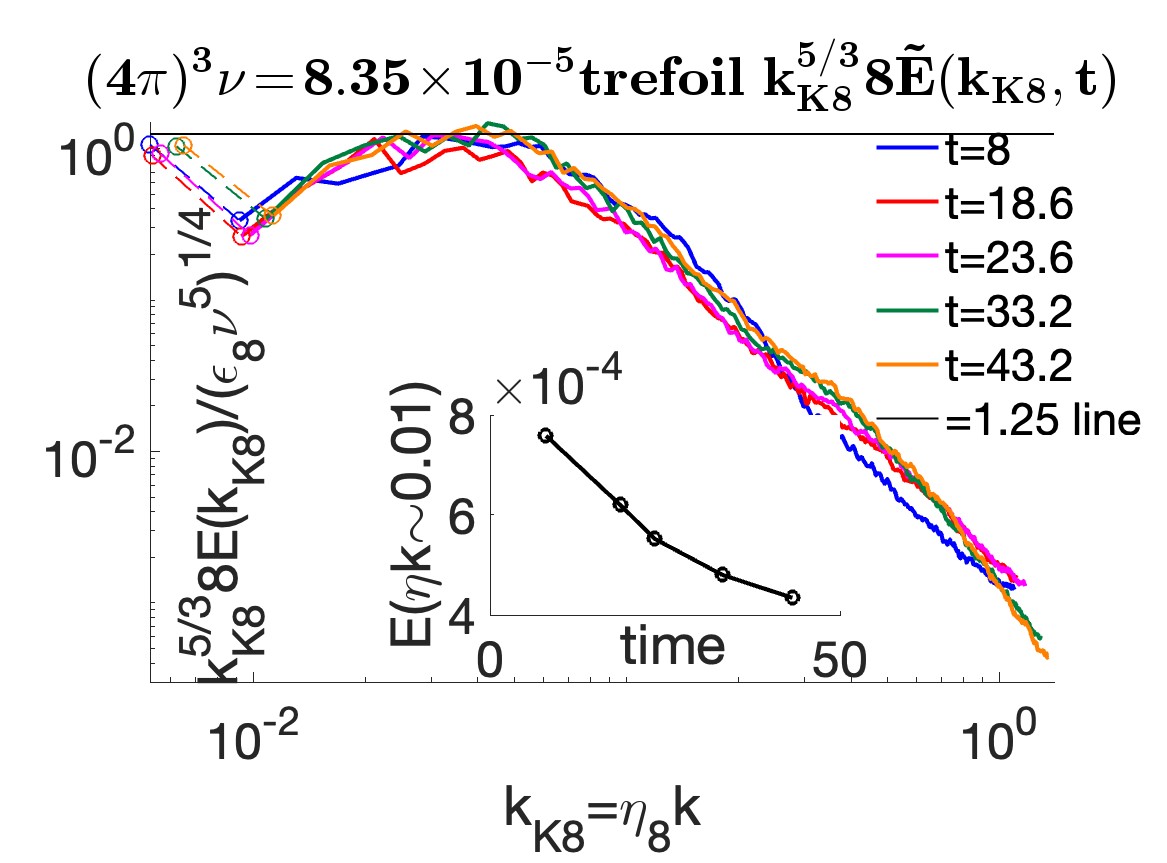}
\begin{picture}
(0,0)\put(-340,86){\large(a)}\put(-140,81){\large(b)} 
\end{picture} 
\caption{\label{fig:r2d054piSpec} Two rescalings of the wavenumber spectra and wavenumbers.
Note first that both $E(k)$ and $\epsilon(t)$ are rescaled by a factor of 8 to 
$E_8(k)=8E(k)$ and $\epsilon_8=8\epsilon$ with $\eta_8=(\nu^3/\epsilon_8)^{1/4}$
because the initial trefoil fits within the inner $(2\pi)^3$ domain, and past 
numerical experience going from \cite{YeungetalJFM2025} back to \cite{KerrJFM1985} 
comes from forced HIT calculations in $(2\pi)^3$ domains
Second, both frames use Kolmogorov scaled $\eta_8 k$ wavenumbers
and $(\epsilon_8\nu^5)^{1/4}$ scaled $E(k)$. 
The differences are:
(a) The $E_8(k)$ spectra are not multiplied by $(\eta_8)^{5/3}$ so as to show 
the similarities to the 
new experimental spectra of \cite{MatsuetalPNAS2026}.  Note that without rescaling,
all of the spectra decay with time.
(b) The same data but multiplied by $(\eta_8k)^{5/3}$ and including the $k\sim0$ shell. 
A plausible Kolmogorov constant of 1.25 runs through the spectra.
The rescaled $(\eta_8k)^{5/3}E_8(k)\leq0.005)$ is relatively large, with the inset
showing how $E(\eta k\!\sim\!0.01)$ (and all greater $k$) are decaying with time.}
\end{figure}

\smallskip
To replicate these results numerically, this paper extends in time and domain size a modest Reynolds number trefoil vortex knot calculation \citep{KerrPRF2023} using the more compact Rosenhead  algebraic vorticity profile. Initialisation is done
in the larger $(4\pi)^3$ domain by putting the original initial condition 
into a $(2\pi)^3$ subdomain, applying a hyperviscous filter, and  then remeshing 
onto $1024^3$ resolution. 
Figure \ref{fig:r2d054piSpec} presents the new numerical  spectra in two ways. For comparison with the new experiment \citep{MatsuetalPNAS2026}, the first illustrates the formation of one decade of a $E(k,t)\sim k^{-5/3}$ spectrum that persists for roughly four eddy turnover times. The second demonstrates Kolmogorov similarity.
\begin{figure}
\includegraphics[scale=.30,clip=true,trim=0 180 0 0]{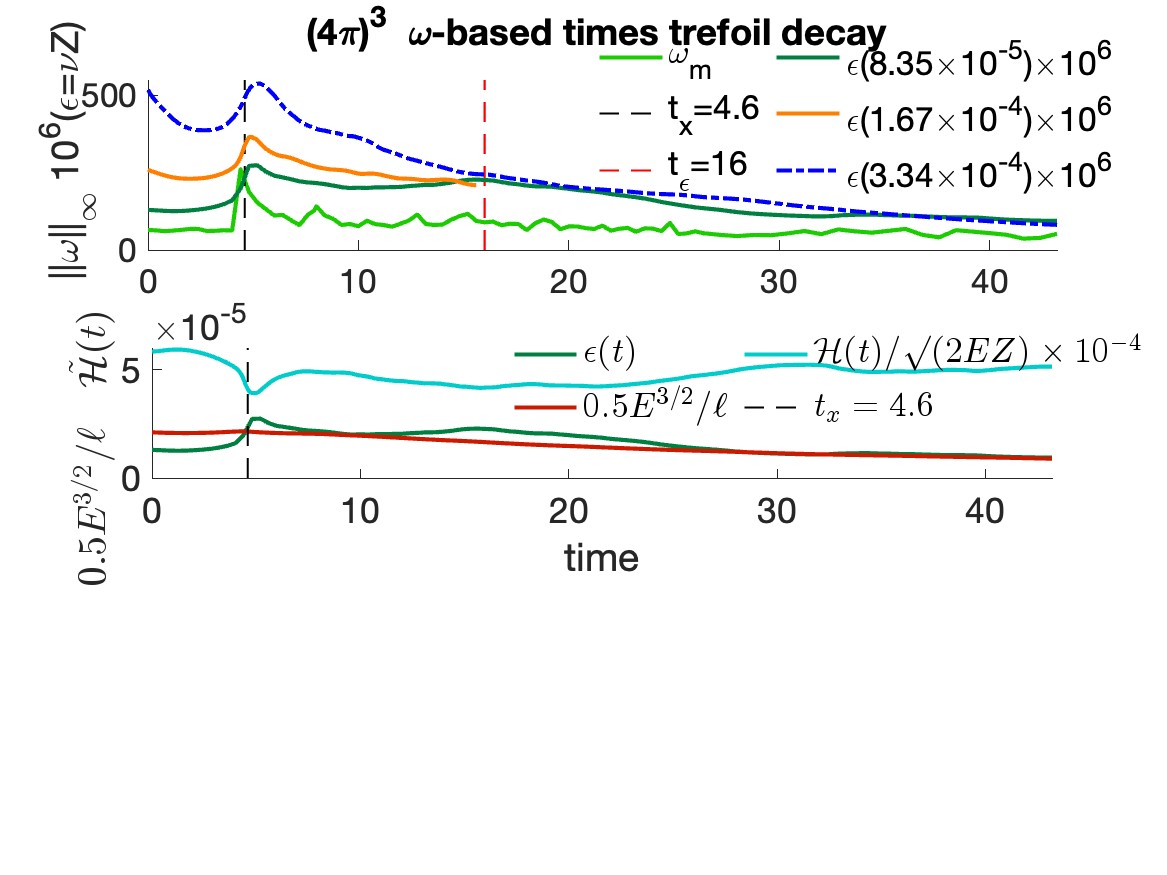} //
\bminic{1}
\vspace{-12mm}
\includegraphics[scale=.28,clip=true,trim=0 400 0 0]{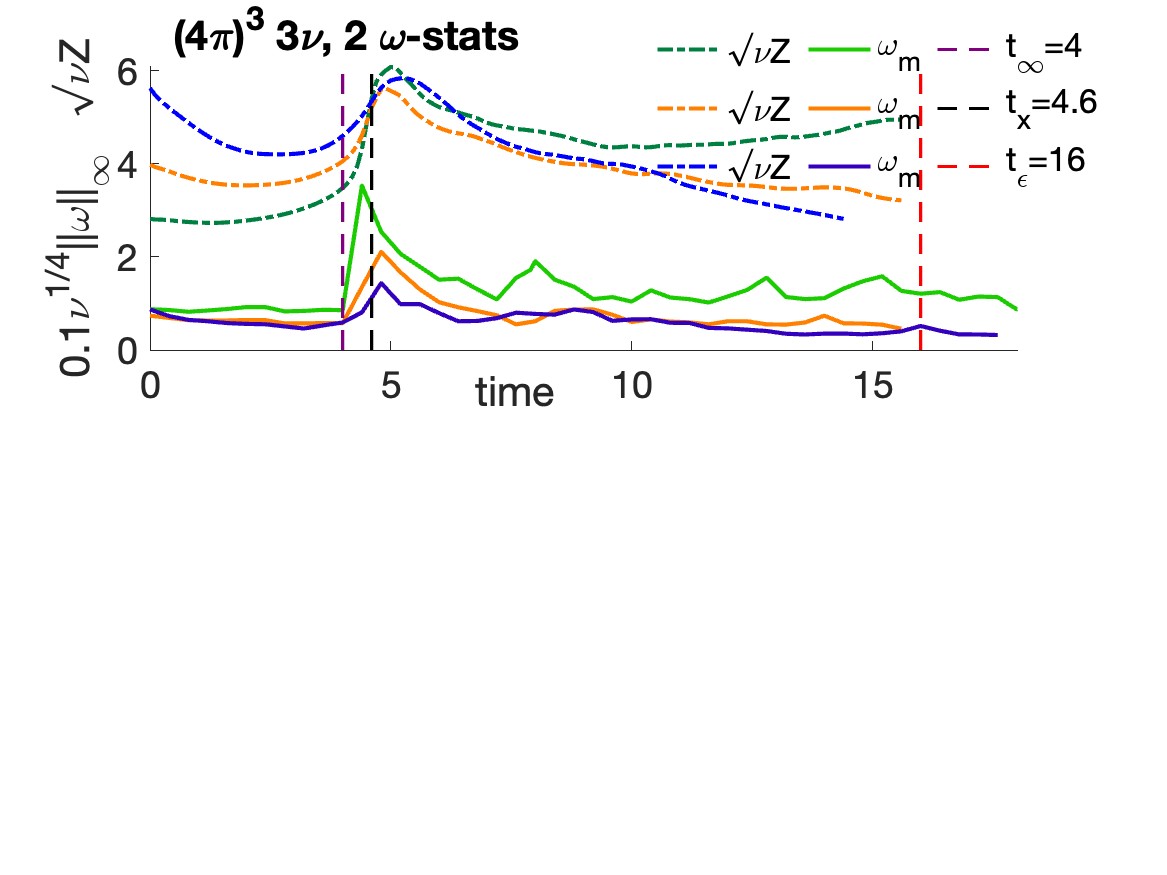}
\emini
\begin{picture}(0,0)\put(-234,231){\Large(a)}
\put(-238,148){\Large(b)} 
\put(-218,62){\Large(c)} \end{picture} 
\caption{\label{fig:KSRZHr2d05sqnu} Global diagnostics designed to show continuity with the $(2\pi)^3$ Rosenhead calculations in \cite{KerrPRF2023}. 
(a) The $\nu=8.35\times10^{-5}$ evolution of the maximum vorticity 
$\omega_\infty(t)=\|\omega(t)\|_\infty$, then the dissipation rates $\epsilon(t)=\nu Z(t)$ 
for $\nu=8.35\times10^{-5},~1.667\times10^{-4},~3.334\times10^{-4}$
($Z$ is the enstrophy \eqref{eq:enstrophy}).
The spike in $\omega_\infty$ at $t=4$ is followed by the maxima of the dissipation 
rates for all at $t\sim0.5$, as found for 
the $(2\pi)^3$ calculation \citep{KerrPRF2023}, 
and represents roughly when the first reconnection ends. 
(b) For $\nu=8.35\times10^{-5}$, $\epsilon(t)$, 
a $\tilde{\epsilon}=0.5E^{3/2}/\ell$ estimate, and
a scaled  helicity $\cH(t)/\sqrt{2E(t)Z(t)}$.  
After $t\geq15$, the decaying $\epsilon(t)$ and the scaled 
helicity both settle down. That the post-reconnection scaled helicity is relatively 
constant, suggests the preservation of some aspect of the global topological structure.
(c) For all $\nu$, early time $\nu$-independent convergences of 
$\nu^{1/4}\|\omega(t)\|_\infty$ and $\sqrt{\nu}Z(t)=(\nu^{1/4}\Omega_{V1}(t))^2$ at 
$t_\infty=4$ and $t_=t_x=4.6$ respectively, with $t_x$ showing when the first 
reconnection ends. Not shown is the full $\nu^{1/4}\Omega_{Vm}(t)$ hierarchy with 
convergences at their $t_m$, as in \cite{KerrJFM2025}.
}
\end{figure}

The numerical method is based upon earlier vortex knot simulations 
\citep{KerrJFMR2018b,KerrJFMR2018c} that replicated the topological changes of the 
series innovative vortex knot experiments 
\citep{ScheeleretalIrvine2014,Scheeleretal2017} that has now led to the more recent 
experiment by \citet{MatsuetalPNAS2026}. 
Trefoil knot simulations using algebraic vorticity profiles have several 
mathematical advantages. First, they are compact in the sense that their velocity \eqref{eq:NS} and vorticity \eqref{eq:omega} decay exponentially from their centres, thus minimizing the influence of  the periodic boundaries.  A trefoil also has strong, non-trivial helicity due to its topology.
\par\smallskip
However, a disadvantage of using trefoil knots comes from how the vorticity of the strongly curved vortices is mapped onto a Cartesian mesh. Using recent stability analysis \citep{GallaySmets2020} and simulations, \citet{KerrPRF2023} has shown why algebraic profiles are a better choice than Gaussian profiles, mainly because they 
allow the shedding of finer scale structures such as $h<0$ helical vortex 
sheets, whose primary reconnection-scale properties have 
now been identified for several algebraic profile simulations 
\citep{KerrJFMR2018b,KerrJFMR2018c,KerrPRF2023,KerrJFM2025}. 
Figures \ref{fig:3Dstruc}a,d show the sheets' remnants and their primary 
associated properties include
temporally local scaling with a set of $\nu$-independent convergences of the 
higher-order vorticity moments \citep{KerrJFM2025}, 
as illustrated in figure \ref{fig:KSRZHr2d05sqnu}c, with the sheets'
post-reconnection interactions generating finite energy dissipation rates 
and eventually temporally-integrated finite dissipation as 
in \ref{fig:KSRZHr2d05sqnu}a. 
\par\smallskip
This paper focuses more on the evolution after the first reconnection and is organized as follows. The equations and numerics are outlined in
§\ref{sec:equations} followed by the definition of the trefoil vortex knot.  
After that, in §\ref{sec:Kspectra} the spectral forms are plotted  in figure \ref{fig:r2d054piSpec}. Then §\ref{sec:global} discusses the evolution of  several global vorticity, energy and helicity measures that connect these results to recent work \citep{KerrPRF2023,KerrJFM2025}, and which also provide insight into the times chosen for the three-dimensional structures in figure \ref{fig:3Dstruc}. The first of these times appears just before the trefoil separates into two rings, before evolving into two triangular rings at a later time. §\ref{sec:lengths} contains a discussion of how the large scale diagnostics grow. 

\subsection{Governing equations \label{sec:equations}}

The governing equations are the three-dimensional, incompressible Navier-Stokes equations for the velocity ($\mbox{div}\, \bu = 0$) and the vorticity $\bomega=\nabla\times\bu$.
\EQL{eq:NS} \hspace{-8mm} 
\ppto{\bu} + ({\bu}\cdot\nabla){\bu} = -\nabla p+ 
\nu\triangle{\bu}\,,
\EN
\EQL{eq:omega} 
\hspace{-8mm} \ppto{\bomega} + ({\bu}\cdot\nabla){\bomega} = 
({\bomega}\cdot\nabla){\bu} + \nu\triangle{\bomega}\,.
\EN
These equations have three quadratic diagnostics, the continuum densities 
of the energy $e=\half|\bu|^2$, enstrophy $\zeta=|\bomega|^2$ and
helicity $h=\bu\cdot\bomega$, which can take either sign. 
The equations for $e$ and $\zeta$, plus their integrated values, are
\EQL{eq:energy}
 \hspace{-15mm} \ppto{e}+ ({\bu}\cdot\nabla)e = -\nabla\cdot(\bu p) 
+\nu\triangle e -\nu(\nabla\bu)^2,\qquad 
E=\half\int\bu^2dV\,,
\EN
\EQL{eq:enstrophy} 
\hspace{-15mm} \ppto{\zeta}+ ({\bu}\cdot\nabla)|\bomega|^2 = 
2\bomega\bS\bomega +\nu\triangle|\bomega|^2
-  2\nu(\nabla\bomega)^2\,, \qquad Z=\int\bomega^2dV\,,
\EN
where the local dissipation rates are $\epsilon=\nu\zeta=\nu(\nabla\bu)^2$ and $\epsilon_\zeta=2\nu(\nabla\bomega)^2$ and the enstrophy production rate is
$\zeta_p=\bomega\cdot\bS\bomega$.
\par\smallskip
Due to these robust relationships between the energy density $e$  and enstrophy 
density $\zeta$, Reynolds number-independent energy dissipation, 
a {\it dissipation anomaly}, $\Delta E_\epsilon=\int_0^{T_\epsilon}\epsilon\,dt>0$, 
can form in a finite-time $T_\epsilon$ at large Reynolds numbers for another
set of algebraic profile trefoils \citep{KerrJFM2025}.  
Negative $h<0$ vortex sheets play a crucial role in generating finite 
$\Delta E_\epsilon$ \citep{KerrPRF2023,KerrJFM2025}, with the equation for 
$h$, and the integrated helicity $\cH =\int\bu\cdot\bomega dV$, obeying
\EQL{eq:helicity} 
\hspace{-15mm} \ppto{h}+ ({\bu}\cdot\nabla)h =\!
\underbrace{-\bomega\cdot\nabla\Pi}_{h_f=\omega-{\rm transport}}
\!+\!\underbrace{\nu\btriangle h}_{\nu-{\rm transport}}\!-\underbrace{
2\nu{\rm tr}(\nabla\bomega\cdot\nabla\bu^T)}_{\epsilon_h={\cal H}-{\rm dissipation}}\,.
\EN
Another set of diagnostics associated with the vortex sheets are 
higher-order vorticity moments 
$\Omega_{Vm}(t)=(\int_{V\ell}|\omega|^{2m}dV)^{1/2m}$. Those moments are being
volume-integrated to allow comparisons between calculations done in 
different periodic domains \citep{KerrJFM2025}, a task made easier here because 
all the new calculations use a $V=(4\pi)^3$ domain. 
Discussed next is one way to address similar variable domain issues when comparing 
spectra from simulations performed in different domains. 
\par\smallskip
Once a means of comparing variable domains was determined, a set of 
$\nu$-independent, $t_m$ temporal convergences was found by taking 
$\nu^{1/4}\Omega_{Vm}(t)$, with this hierarchy of the $t_m$:  
$t_\infty\leq\dots t_m\dots\leq t_1=t_x$. Figure \ref{fig:KSRZHr2d05sqnu}c shows
that scaling for $\nu^{1/4}\|\omega\|_\infty$ and 
$\sqrt{\nu}Z=(\nu^{1/4}\Omega_{V1})^2$ for the Rosenhead profile calculations,
with all the remaining scaled moments sandwiched between those. The origin of this
new $\nu^{1/4}$ scaling is being investigated using a single orthogonal vortex pair
with Rosenhead vorticity profiles.
\par\smallskip
The simulated viscosities are\,: $\nu=8.35\times10^{-5}$ and $\nu=1.667\times10^{-4}$ on $n^3=1024^3$ periodic meshes
and $\nu=3.334\times10^{-4}$ on a $512^3$ mesh.  Following \cite{BustaKerr2008}, the 
calculations are 2/3rds dealiased with a hyperviscous filter.
\par\smallskip
The vortex knot is initialised as in \cite{KerrPRF2023} with the two primary elements 
being the trajectory of the trefoil centreline and the mapping of the vorticity 
profile $\omega_{\footnotesize\mbox{raw}}(\rho)$ onto the mesh using the distance $\rho$
and tangent $\partial\bxi_0(\phi)$ of the nearest point on the trajectory.
The initial closed, unperturbed, double loop trajectory of all of the trefoils is
$\bxi_0(\phi)=[x(\phi),y(\phi),z(\phi)]$ $=[r(\phi)\cos(\alpha),r(\phi)\sin(\alpha),a\cos(\alpha)]$
with $\phi$=1:4$\pi$ and $a=0.5$, $w=1.5$, $r_f=1$,
$r(\phi)=r_f+a\sin(w\phi)$ and $\alpha=\phi+a\cos(w\phi)/(wr_f)$.
The algebraic profiles $\omega_{\footnotesize\mbox{raw}}(\rho)$ are parameterized 
by a radius $r_\circ$, its maximum/centerline vorticity $\omega_\circ$ and
a power $p_r$, giving  $\omega_{\footnotesize\mbox{raw}}(\rho)=
\omega_\circ(r_\circ^2)^{p_r}/(\rho^2+r_\circ^2)^{p_r}$, followed by a 
hyper-viscous smoothing.
For the algebraic Rosenhead profile here, $p_r=2$ and $r_\circ=0.05$. 
\par\smallskip


\section{Results: Three-dimensional spectra and temporal scaling \label{sec:Kspectra}}

Energy spectra $E(k,t)$ are formed by taking discrete Fourier transforms 
and for the $L^3=(4\pi)^3$ domain with $n=1024$, the minimum wavenumber shell is 
$k_{\rm min}=\min_{k\neq0}(k)=2\pi/L=1/2$ and when moderated by the $\eta$ microscale
and plotted, $\min(\eta k)\sim0.01$. With the wavenumber 
shells $E(k,t)$ filled using\,: 
\EQL{eq:Ekt} 
E(k,t)=\frac{1}{2}\sum_{k}|\bu_{\bk}(t)|^2 \qquad \text{over}\qquad
k\leq|\bk|<k+k_{\rm min}\,.
\EN 
Historically, attempts to generate persistent $E(k,t)\sim k^{-5/3}$ spectra from  
unforced decaying turbulence have used only a small set of particularly unstable
Fourier modes, such as the Taylor-Green vortex \citep{BrachetetalJFM1983}, or random 
initial conditions with a spectral peak at small or moderate wavenumbers
\citep{IshidaDavidsonKJFM2006,RodhiyaSreeniPTRSA2026}. For both, there is an inherent
limitation when the developing turbulence reaches the periodic boundaries, 
reacts back upon itself, and inhibits the formation of -5/3 spectral cascade. 
This raises two questions. 
\begin{enumerate}
\item~~Could an unstable vortical configuration do better? \cite{KerrJFM2025} showed that a {\it dissipation anomaly} \eqref{eq:energy} is possible if the periodic domain increases as the Reynolds number increased. However, the extent of the -5/3 regime was limited by the broad algebraic profile that was used. 

\item~And could the more compact algebraic Rosenhead profile from \cite{KerrPRF2023} do even better? 
Possibly, but only in a larger $(4\pi)^3$ domain. The calculations in this paper contribute to that interpretation.
\end{enumerate}
What is the effect of domain size?
Because the initial condition was originally created for a $(2\pi)^3$ domain, 
the initial condition is very inhomogeneous and 
the volume-averaged spectra are a factor of $2^3=8$ smaller 
than those of the original $(2\pi)^3$ calculation or 
forced calculations done in $(2\pi)^3$ domains. To compensate, all 
the energies and enstrophies used for scaling the spectra are multiplied 
by $8=2^3$. The result in figure \ref{fig:r2d054piSpec} is spectra
that are similar to those from the earlier $(2\pi)^3$ 
calculation \citep{KerrPRF2023} and spectra from homogeneous turbulent flows, 
including forced homogeneous, isotropic (HIT) calculations in $(2\pi)^3$ 
domains going from \cite{YeungetalJFM2025} back to \cite{KerrJFM1985}.
\par\smallskip
With that background, figure \ref{fig:r2d054piSpec} presents two rescalings 
of the $E(k,t)$ spectra, with both rescalings sharing Kolmogorov scaled 
wavenumbers $k_K=\eta_8(t)k$, with $\eta_8=(\nu^3/(8\epsilon(t)))^{1/4}$,
and with the $8E(k,t)$ scaled by $(8\epsilon(t)\nu^5)^{1/4}$. With these differences\,:
\par\smallskip
\ITM
\item[(a)] For qualitative comparisons with the similarly-scaled 
experimental spectra in \cite{MatsuetalPNAS2026}, the 
$E(k_{K8},t)/(8\epsilon(t)\nu^5)^{1/4}$ spectra, without $E(\min_k)$, are plotted.
\item[(b)] To provide more quantitative comparisons with HIT simulations and 
the $(2\pi)^3$ spectra for the same initial condition, the $\min_k$ shells 
are included and the spectra are multiplied by $k_K^{5/3}$ to provide an
estimate for the Kolmogorov constant $C_K$ if $8E(k_{K8},t)$ obeys
\ITN
\EQL{eq:K41} 8E(k_{K8},t)\sim C_K 
\epsilon_8^{2/3}k_{K8}^{-5/3}\quad\text{with}\quad C_K\sim1.25\,.\EN

\subsection{Evolution of three-dimensional diagnostics and structure \label{sec:global}}

The focus of this paper has been upon the period between when the first reconnection ends at
$t_x\approx4-4.6$, and $t_f\sim40\gtrsim 7t_x$, with Kolmogorov similarity still seen in figure 
\ref{fig:r2d054piSpec} has lasted for the, with figure \ref{fig:KSRZHr2d05sqnu}a showing one of the 
peaks of $\|\omega(t)\|_\infty$ near $t_x$ and strong $\epsilon(t)$ appearing 
for all $t>t_x$. 

\par
\begin{figure}
\bminic{0.55}
\includegraphics[scale=.21]{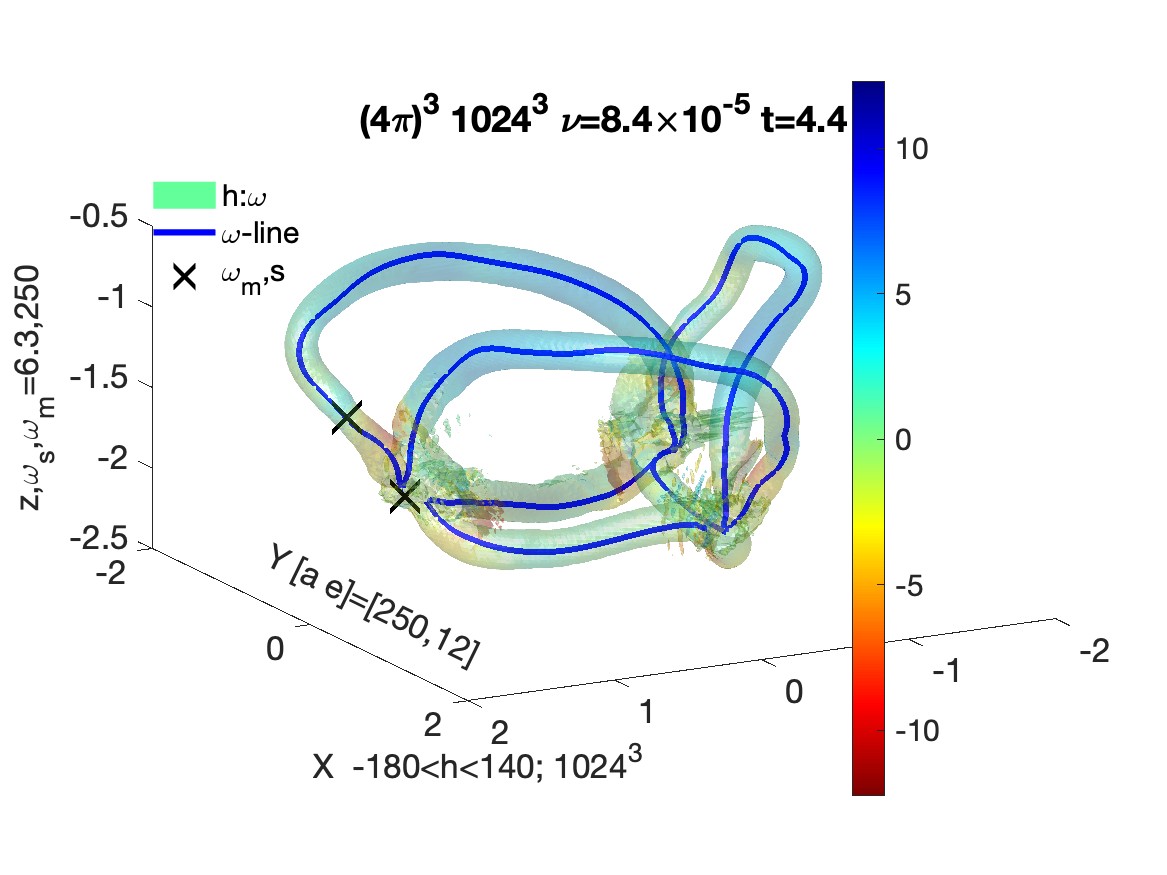} \\
\vspace{-4mm}
\includegraphics[scale=.36,clip=true,trim=0 32 0 0]{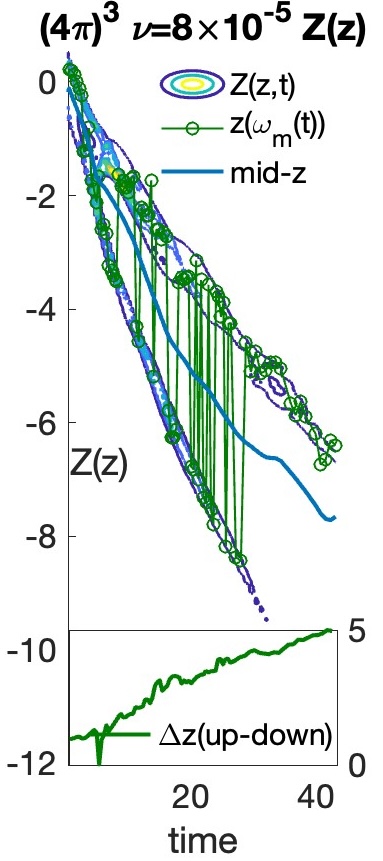}
\emini \bminic{0.45}
\includegraphics[scale=.18,clip=true,trim=0 4 0 56]{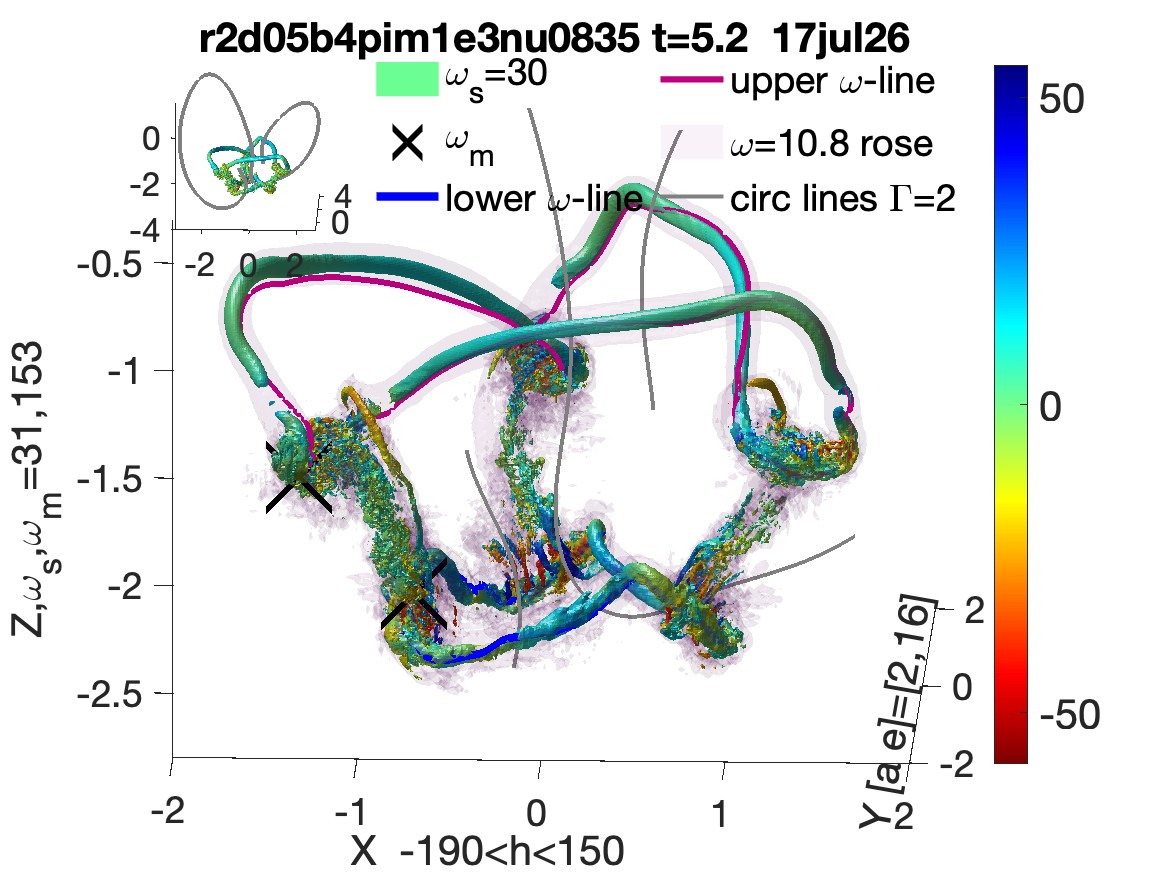} \\ 
\includegraphics[scale=.18,clip=true,trim=0 0 0 60]{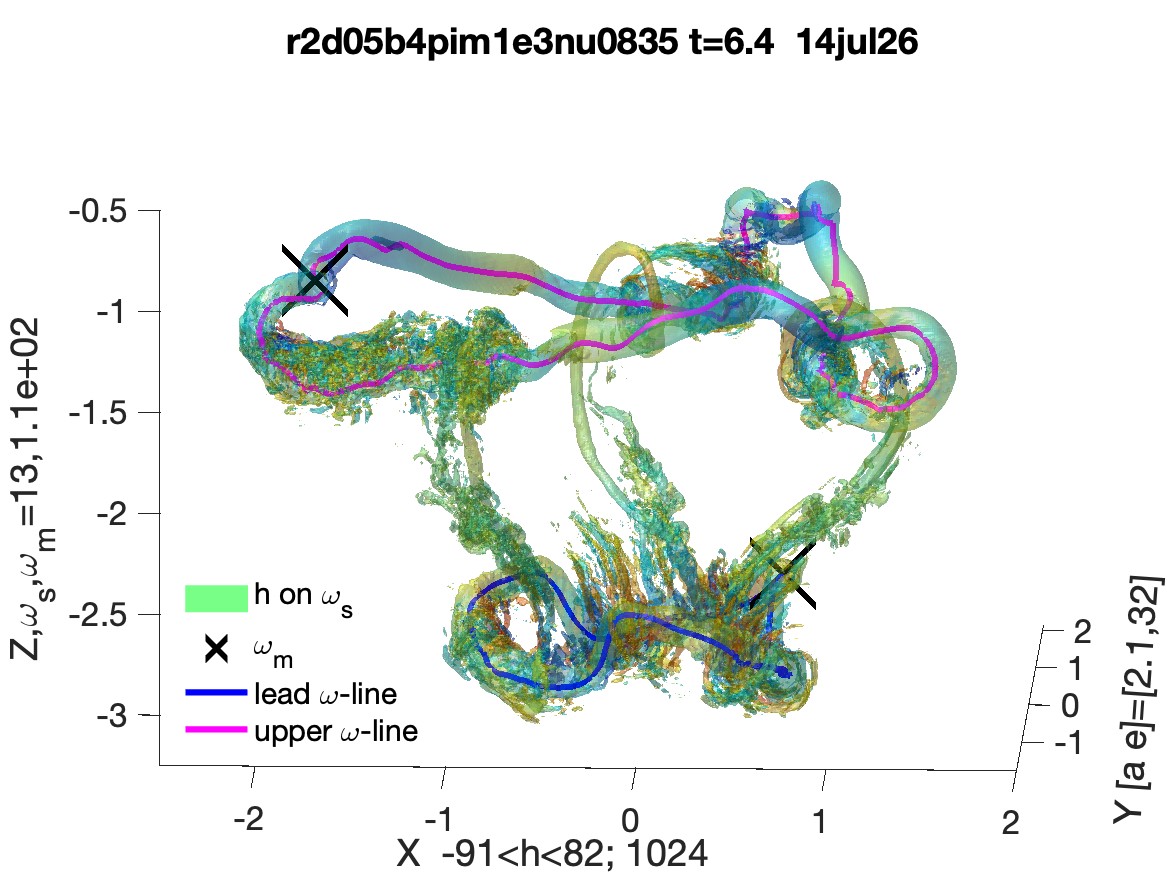}\\~ \\
\includegraphics[scale=.18,clip=true,trim=0 0 0 60]{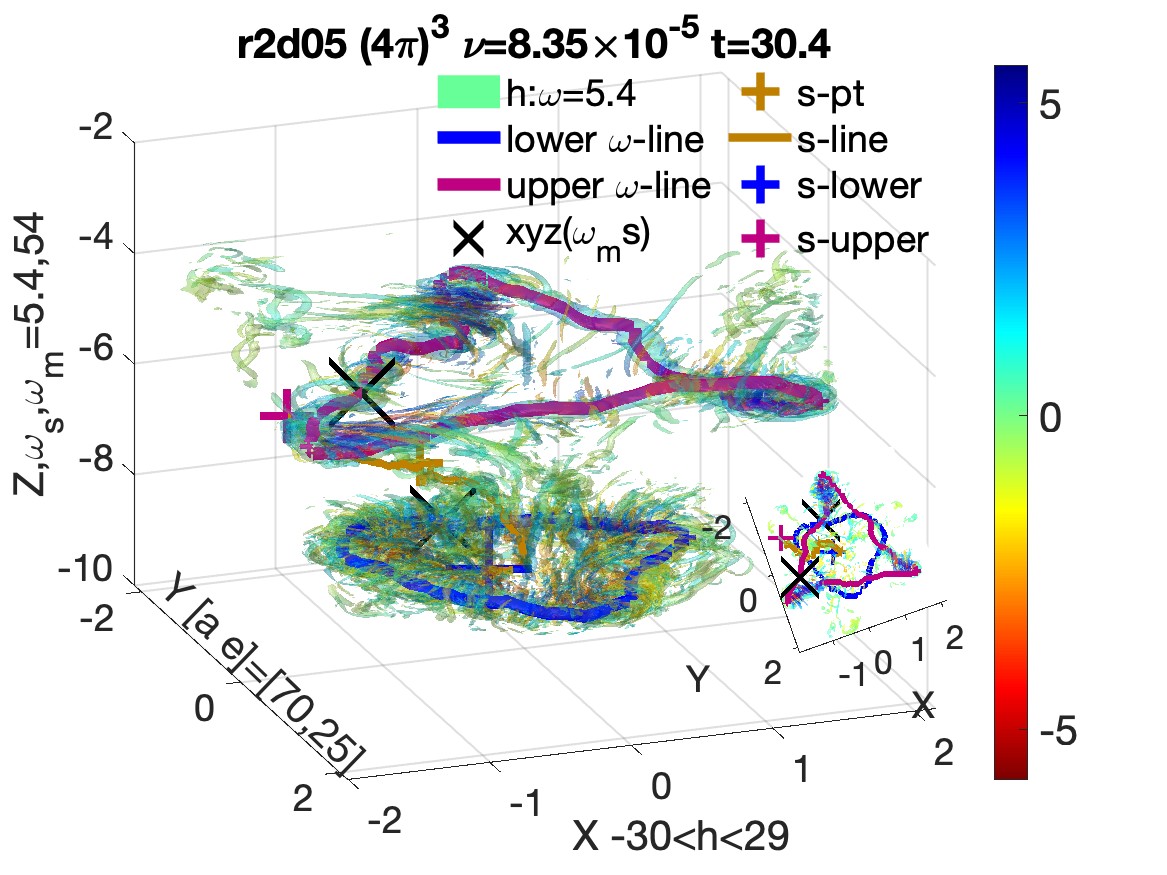} 
\emini
\begin{picture}(0,0)\put(-240,200){\large(a)}\put(-240,0){\large(b)}
\put(-240,-180){\large(c)}  \put(-340,-168){\Large Time}  
\put(-15,160){\large(d)}\put(-75,220){\bf\large t=5.2}
\put(0,0){\large(e)}\put(-85,52){\bf\large t=6.4 }
\put(-15,-140){\large(f)}\put(-85,-78){\bf\large t=30.4}  \end{picture}
\caption{\label{fig:3Dstruc} Splitting of trefoil into two rings is shown 
using three-dimensional $h$-mapped isosurfaces (dominated by green to blue $h\gtrsim0$), local 
$\omega_m$ locations at $X$'s that seed the upper (deep red) and lower (blue) centrelines,
plus vertical profiles.. 
(a) $t=4.4$ shows $|\omega|=6.3$ just before reconnection finishes with only one vortex line..
(b) Vertical profiles of the enstrophy $Z(z,t)$ showing that the vertical separation of the 
new rings begins at $t\approx5$, as the two rings propagate to $-z$, with $z$-mid the
average $z$ between the two rings.
The $z(\omega_m(t))$ line charts how the maximum of vorticity jumps between 
the upper and lower rings. 
(c) This charts the separation between the two rings, which provides one mechanism 
and measure for how the large scales increase. $\Delta z=0$ only at $t=5$.
Frames d-f follow the $\omega$-rings as both are propagating to 
$-z$. (d) $t=5.2$, just after the first reconnection with an additional rose $\omega=10.8$ 
isosurface to represent a $h\sim-10$ remnant sheet, through which $h=\pm20$ braids poke. Plus
an inset showing two closed grey $\bu$-streamlnes. (e) $t=6.4$ is similar to $t=5.2$ with
additional braids that connect the rings.
(f) $t=30.4$, with with the inset showing the rings final, off-set triangular shapes.
A yellow curve in both projections shows how the two rings are connected.}
\end{figure}

\smallskip
Figure \ref{fig:KSRZHr2d05sqnu}b shows two ratios that underlie aspects of the $t>16$ 
dynamics. The first is $\cH(t)/\sqrt{2E(t)Z(t)}$, which is surprisingly constant 
for $t>7$, despite the significant change in $\cH(t)$ \eqref{eq:helicity} at 
$t\sim t_x=4.6$ due to reconnection. This steady $\cH(t)/\sqrt{2E(t)Z(t)}$ might be due to 
an underlying, persistent structure between the new rings in figure \ref{fig:3Dstruc} 
as they propagate downwards.
\par
\begin{figure}
\includegraphics[scale=.32,clip=true,trim=0 0 0 0]{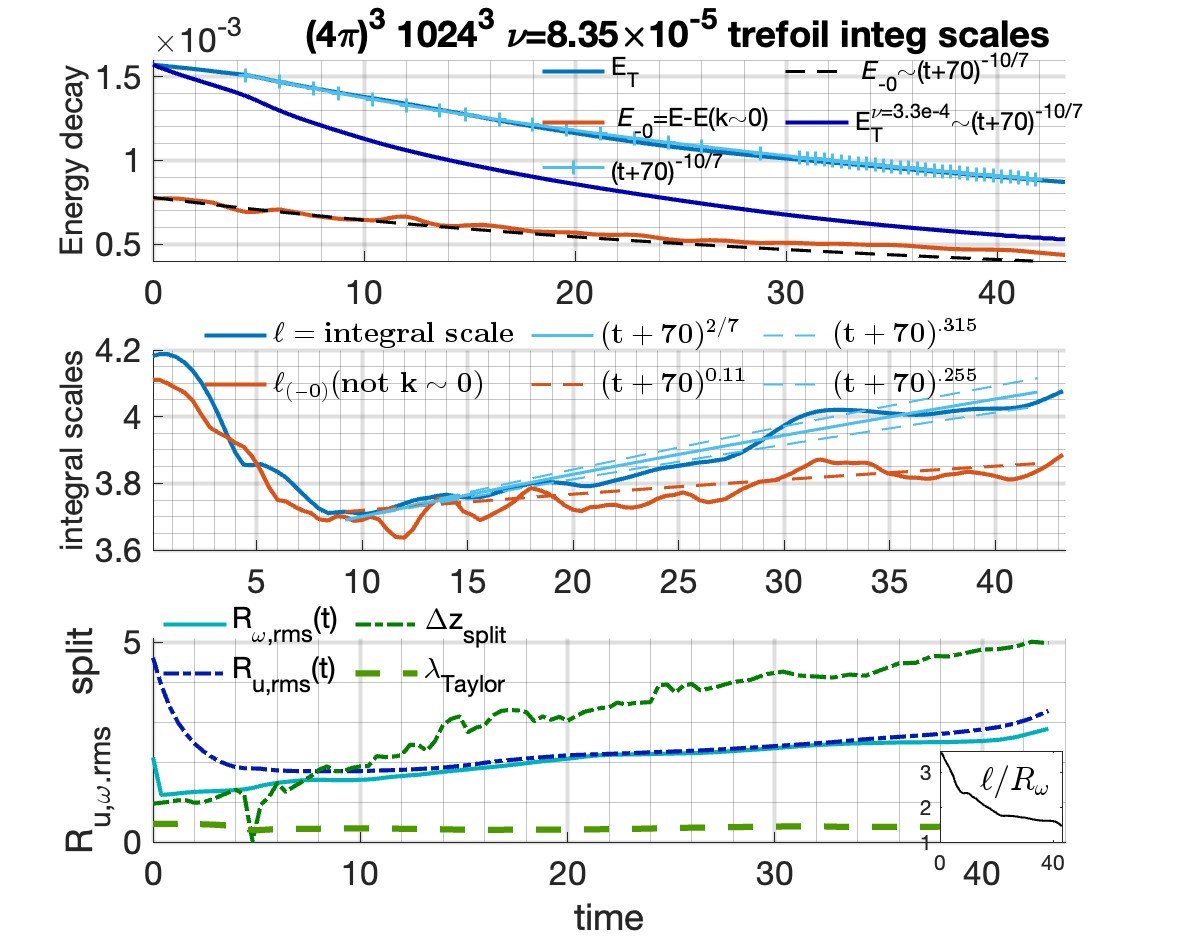}
\begin{picture}(0,0)\put(-310,255){\Large(a)}\put(-280,160){\Large(b)} 
\put(-205,85){\Large(c)}\put(-78,38){\large(d)} \end{picture} 
\caption{\label{fig:r2d05EtINTG} Time-dependencies of Rosenhead profile energy and 
velocity-based length scales. Most use $\nu=8.35\times10^{-5}$. 
(a) Two measures of the energy decay.  Plausible fits for all are $E\sim(t+70)^{-10/7}$, 
which would be consistent with the scaling of based upon the Loitsyanskii integral 
\eqref{eq:Ekk}.  $E_T(t)$ is the full energy, with 
$E_T^{\nu=3.3\times10^{-4}}\sim(t+70)^{-10/7}$ obeyed to t=92. 
${\it E}_{-0}=E-E(k=1)$ does not include the $\eta k\sim0.01$ Fourier modes, for which
another plausible, but less exact, fit would be $(t+60)^{-6/5}$, the Saffman
$\cL$ prediction and what is observed for grid decay 
\citep{KrogstadDaJFM2010,SinhuberetalPRL2015}. 
(b) Two measures of the large, integral length scale $\ell$ \eqref{eq:intL}, one for
$E_T(t)$ with $\ell\sim t^{2/7\pm0.03}$, the Loitsyanskii-based scaling, and 
the other for ${\it E}_{-0}$, 
$\ell_{(-0)}\sim(t+70)^{0.11}$ (dashed-red-orange), which is the same order as the
$\ell\sim t^{0.16}$ scaling of the new experiment \citep{MatsuetalPNAS2026}.
Neither is near the $\ell\sim t^{2/5}$ Saffman prediction \citep{KrogstadDaJFM2010}. 
(c) Root-mean-square radii based upon the energy and enstrophy, which are both 
controlled by the horizontal sizes/radii of the two rings. Plus 
the vertical separation into two swirling rings and the
Taylor microscale $\lambda\leq0.5$,
with $\max\left(R_{8\lambda}(t\geq5)\right)\approx500$.
(d) Inset the ratio of $\ell$ and $r_{\omega,\rm{rms}}$.
}
\end{figure}

\smallskip
The second is a classical turbulence estimate of the dissipation rate 
$\tilde{\epsilon}=0.5E^{3/2}/\ell$ that uses the integral scale $\ell$ \eqref{eq:intL}.
This is compared to the directly determined dissipation rate $\epsilon(t)$ 
\eqref{eq:energy} in figure \ref{fig:KSRZHr2d05sqnu}b. That relatively good comparison 
would support the predictions for energy decay and integral scale growth based upon 
the Loitsyanskii integral \eqref{eq:Ekk} and given in figure \ref{fig:r2d05EtINTG}a,b. 
\par\smallskip
Figure \ref{fig:KSRZHr2d05sqnu}c shows, for the three viscosities, 
the $t\to t_m$ convergences of two vorticity diagnostics. For $m=1$, or
equivalently for the reconnection-enstrophy 
$\sqrt{\nu}Z(t)=\left((\nu^{1/4}\Omega_{V1}\right)^2$ (after \eqref{eq:enstrophy}) and
$m=\infty$, $\nu^{1/4}\|\omega\|_\infty$. These are
at $t_x=t_1=4.6$ and $t_\infty=4$ respectively. Additional $\Omega_{Vm}$ have been
calculated, with all of the $t_m$ convergence times sandwiched between $t_\infty=4$
and $t_x=4.6$, as in \cite{KerrPRF2023}.
\par\smallskip
The evolution of the three-dimensional vorticity in figure \ref{fig:3Dstruc} uses 
helicity density, $h$-mapped, vorticity isosurfaces. Using $h$ has previously helped in 
identifying the dynamical origins of the smallest scale vortex structures during 
reconnection, including showing how the $h_f=\omega$-transport term
\eqref{eq:helicity}) leads to the $h\leq 0$ vortex sheets (Kerr 2023). As before
\citep{KerrPRF2023,KerrJFM2025}, those sheets allow the global enstrophy $Z(t)$ 
and the energy dissipation rates $\epsilon(t)$ to continue increasing after 
reconnection. An event that transforms the trefoil into two connected vortex rings, 
as illustrated using two types of visualisation in figure \ref{fig:3Dstruc}, and
observed in experiments \citep{KlecknerIrvine2013}. This
begins at $t=4.4\lesssim t_x=4.6$ in figure \ref{fig:3Dstruc}a when only one 
centerline vortex can be identified, along with reddish and yellow $h\lesssim0$ 
sheets that are still being shed of off the 
three self-reconnecting local knots at the bottom. 
\par\smallskip

The evolution of the new vortex rings from $t=5.2$ onwards is given in
figures \ref{fig:3Dstruc}d-f. Also included in these frames are
the post-reconnection vortex structures the develop out of the $t<5.2$
vortex sheets and new structures being shed by the rings. With
one goal of these visualisations being to provide suggestions for how the $k^{-5/3}$ spectra 
and strong energy dissipation rates form.

At $t=5.2$, the primary $\omega=30$ isosurfaces are shown in tones of $h\gtrsim0$ 
green that can be divided into three zones. First, the upper vortex ring with 
a red centreline.  Second, the lower vortex ring with a blue centreline. And third, 
three braided bridges that are embedded within a second $|\bomega|=10.8$, $h<0$ 
rose isosurface that is a remnant of the earlier $h\lesssim0$ vortex sheets. 
Where these bridges meet the rings are where the two rings had formed out of the 
trefoil.  And near the bottom of one bridge in particular, new alternating 
red/blue ($h\pm0$) braids are poking through the rose-coloured sheets before 
extending upwards.
\par\smallskip
Two additional grey velocity $u$-streamlines are also shown, partially in the
main figure and more fully in the inset that shows that these streamlines are 
closed $\Gamma$-rings. Once identified, $\Gamma=\oint \bu\cdot d\bs$ integrals 
about the two 
rings generate circulations of $\Gamma\approx2$ (for the trefoil's two loops). 
This property is preserved from $t=0$ to $t\sim10$ and could be a source 
for the Loitsyanskii integral $\cI$ \eqref{eq:Ekk}. However, for $t>10$ the 
closed loops want to (but cannot) extend beyond the periodic boundaries, 
making this method for calculating $\Gamma$ untenable.
\par\smallskip
Frame \ref{fig:3Dstruc}e at $t=6.4$ shows how the tiny ($h=\pm20$) 
t=5.2 braids evolve into larger transient structures between the rings, 
structures that over time are being 
continually renewed, stretched, then coiled around those rings. 
Frame \ref{fig:3Dstruc}f shows that this is a 
recurring pattern that could provide a mechanism for a continual 
refreshment of the $k^{-5/3}$ spectra that persist over these times. 
Note that over time, each ring becomes triangular with 
a relative rotation of $60^\circ$ to the other, as shown by the inset. 
The yellow $s$-line shows the path of one of those braids between the rings, beginning 
at the red $+$ on the upper ring and ending at the blue $+$ in the centre of the 
lower ring (also in inset).
\par\smallskip
Frame \ref{fig:3Dstruc}b quantifies the separation of the rings spawned by
the original trefoil by showing the 
temporal evolution of vertical profiles of the enstrophy $Z(z,t)$, also seen experimentally
\citep{KlecknerIrvine2013}, and how the position 
of $\omega_m$ jumps between the rings over time. With
\ref{fig:3Dstruc}c following $\Delta{z}_{\rm split}(t)$, the distance between
the two vertical $Z(z,t)$ maxima after the vertical enstrophy profile splits 
in \ref{fig:3Dstruc}d. This growth rate is preserved from 
$2t_x\approx10$ to the end of the calculation at $t=43.2$.

\begin{figure}
\includegraphics[scale=.32,clip=true,trim=0 400 0 0]{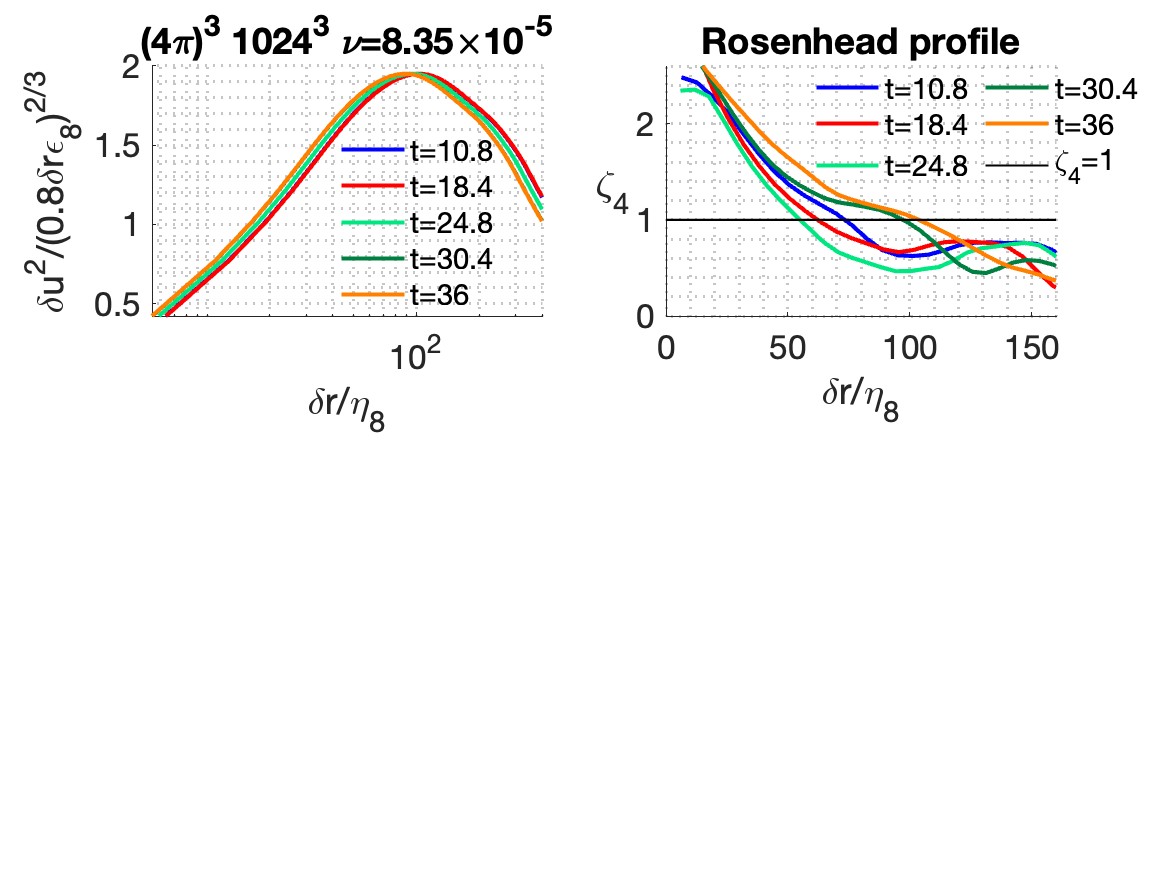}
\begin{picture}(0,0)\put(-210,255){\Large(a)}\put(-80,160){\Large(b)} \end{picture}
\caption{\label{fig:r2d05S2S4} Second and fourth-order velocity structures for t=10.8 to 36.
(a) For $D_{LL}(r,t)=\delta u^2(\delta{r},t)$ and scaled
by the Kolmogorov prediction $\delta u^2(r)/(0.8\delta{r}\epsilon)^{2/3}$. 
The scaled value is consistent with the experimental value \cite{MatsuetalPNAS2026}.
(b) Exponent $\zeta_4$ of the $Dlog_{\delta{r}}$ derivative of the fourth-order 
longitudinal structure function. Note that for $\delta{r}/\eta_8\lesssim100$ the
plateau with $\zeta_4\sim1.2$ is growing.}
\end{figure}

\subsection{Large-scale lengths and the Loitsyanskii invariant $\cI$
\label{sec:lengths}}

Besides the energy spectra, additional decaying turbulence statistics given 
by \cite{MatsuetalPNAS2026} include $E(t)\sim t^{-10/7}$ energy decay and 
the evolution of three large-scale lengths in figure \ref{fig:r2d05EtINTG}c.  
The integral time-scale $\ell(t)$, plus
$R_{u,rms}$ and $R_{\omega,rms}$ for turbulent blob expansion, with
\EQL{eq:intL} 
\ell(t)\equiv (3\pi/4)\int_0^\infty k^{-1}E(k,t)dk \big/ \int_0^\infty E(k,t)dk\,,
\EN
\EQL{eq:Rdef1} 
R^2_{\rm u,rms}(t)=\!\int E(x,t)r^2 dx / \int E(x,t)dx\,;\,
R^2_{\rm \omega,rms}(t)=\!\int E(x,t)\omega^2 dx / \int E(x,t)dx\,. 
\EN
A thorough discussion of the two theories for scaling decaying turbulence, 
based upon the Saffman $\cL$ and Loitsyanskii $\cI$ integrals, are given 
by \cite{IshidaDavidsonKJFM2006} and \cite{KrogstadDaJFM2010}, respectively 
in the context of initial $k\to0$, $k^n$ simulations and 
grid turbulence. One way to describe the origin of $\cL$ and $\cI$ is to
transform the two-point velocity correlation 
$E(k)=\frac{1}{\pi}\int_0^\infty\langle\bu\cdot\bu^\prime\rangle kr\sin(kr)dt$, then
expand the result in $k$
\EQL{eq:Ekk} E(k\to0)=\frac{\cL k^2}{4\pi^2}+\frac{\cI k^4}{24\pi^2}+\dots
~\text{with}~\cL=\int\langle\bu\cdot\bu^\prime\rangle d\br~\text{and}~
\cI=\int r^2\langle\bu\cdot\bu^\prime\rangle d\br\,,
\EN
For grid turbulence there is a well-defined streamwise momentum that gives $\cL\neq0$, 
which scales as $\overline{u}^2\ell^3$ \citep{Saffman1967} and leads to a prediction 
with $E(t)\sim t^{-6/5}$.
In contrast, without a well-defined linear impulse such that $\cL=0$ 
(here the average $\langle u_z\rangle_{x,y}=0$), 
then
$\cI\sim\overline{u}^2\ell^5$ can be used, which leads to $E(t)\sim t^{-10/7}$. 
On dimensional grounds, $\cI$ could go as the angular momentum squared 
$\langle{\mathbf H^2}\rangle/V$ \citep{LandauLifshitzFM1959}, 
which in a finite periodic domain is difficult to define. 
So instead the circulation $\Gamma$ about the vortex knot, with the same 
dimensions, is considered by integrating on $\Gamma$-loops like those in
figure \ref{fig:3Dstruc}a. 

Figure \ref{fig:r2d05EtINTG} shows two versions of the energy decay $E(t)$ in
\ref{fig:r2d05EtINTG}a, 
with the  corresponding integral scales in \ref{fig:r2d05EtINTG}b and further length
scales in \ref{fig:r2d05EtINTG}c. The energy decay for the full fields in shades of 
blue all obey 
$E_T(t)\sim(t+70)^{-10/7}$ and for $\nu=8.35\times10^{-5}$, the integral scale 
grows as $\ell=(t+70)^{2/7\pm0.03}$ in \ref{fig:r2d05EtINTG}b 
When the $E(k\sim0)$ shell from figure \ref{fig:r2d054piSpec}b is removed, the 
decay of the remaining ${\it E}_{-0}(t)$ (red) still goes as $(t+70)^{-10/7}$,
however its integral scale $\ell_{(-0)}$ (red) roughly obeys
$\ell_{(-0)}\sim(t+70)^{0.11}$, the red-dashed in \ref{fig:r2d05EtINTG}b. 

Figure \ref{fig:r2d05EtINTG}c compares the growth of $\Delta z_{\rm split}$ to the 
root-mean-square radii $R_{\rm u,\omega-rms}$ \eqref{eq:intL} about the vortex knot, 
with \ref{fig:r2d05EtINTG}d showing the ratio $\ell/R_{\omega,{\rm rms}}$.
The inset shows that $\ell$ and $R_{\omega-\rm{rms}}$ are
of the same magnitude and converging in time, with the magnitude and growth of
both much less than that of $\Delta z_{\rm split}$.

Note that the experimetntal $\ell_X\sim t^{0.16\pm0.06}$ \citep{MatsuetalPNAS2026} 
is closer to the $\ell_{(-0)}$ curve than to the full
$\ell=(t+70)^{2/7\pm0.03}$ above, suggesting that the experimental spectral 
analysis might be missing wavenumbers at the scale of their apparatus.

As a check on their decay results, \cite{MatsuetalPNAS2026} compares a modelled 
dissipation rate $\tilde{\epsilon}=E^{3/2}/\ell$ to the true dissipation rate 
$\epsilon=dE/dt$. Assuming that $E\sim t^{-10/7}$, then 
$dE/dt\sim t^{-(17/7\approx2.43)}$ and $\tilde{\epsilon}\sim t^{-2.30}$. 
A difference in exponents that would be consistent with the curves in their 
figure 3a. 
Figure \ref{fig:KSRZHr2d05sqnu}b shows an equivalent $\epsilon$ comparison for 
these 
simulations, with $\epsilon(t)=dE/dt(t)$ and $\tilde{\epsilon}=0.5E^{3/2}/\ell$ 
converging for $t\geq25$.

Figure \ref{fig:r2d05EtINTG}c includes $\Delta{z}_{\rm split}(t)$, the separation 
of the rings, with the two versions of $R_{\rm rms}(t)$ \eqref{eq:intL}.
For $t>10$, the two rms radii grow together and more slowly 
than $\Delta{z}_{\rm split}(t)$. The inset, 
figure \ref{fig:r2d05EtINTG}d, shows that $R_{\rm \omega,rms}(t)$ grows faster 
whilst converging to $\ell$, the integral scale in figure \ref{fig:r2d05EtINTG}b. 
And all are significantly
larger than the Taylor microscale $\lambda$ in \ref{fig:r2d05EtINTG}c. 

\noindent{\bf Structure functions} 
Finally, a few longitudinal structure functions 
$\delta{u}^p(\delta{r})$ have been determined, with scaled second-order 
$D_{LL}=\delta{u}^2$ in figure \ref{fig:r2d05S2S4}a and in \ref{fig:r2d05S2S4}b 
the exponent of its fourth-order $Dlog_{\delta{r}}$ derivative,
$\zeta_4=d(\log(\delta{u}^4)/d\log(\delta{r})$. $D_{LL}$ is consistent with the 
equivalent values from \cite{MatsuetalPNAS2026} and for $\zeta_4$ for
$\delta{r}/\eta_8\lesssim100$, a plateau with $\zeta_4\sim1.2$, similar to
forced isotropic turbulence \citep{GotohFukayama2002} is developing.
Not shown: All third-order
 $\tilde{D}_{LLL}=D_{LLL}/(\delta{r}\epsilon)$ have negative slopes for small 
$\delta{r}$, with widely varying slopes, except for $t=30.4$ which has of subrange 
with relatively steady $\tilde{D}_{LLL}(\delta{r})\sim-1$. 
Sixth-order, has a short $\zeta_6\gtrsim1$ regime, 
plus very large oscillations.

\section{Concluding thoughts \label{sec:concluding}}

The three connected threads from this decaying turbulence calculation, with caveats, are:
\ITM\item 
It has been shown how an unadorned trefoil vortex knot can, 
immediately after self-reconnection, generate a $k^{-5/3}$ energy spectrum that 
persists over several turbulent time-scales whilst obeying Kolmogorov scaling. Further
results include a $E(t)\sim t^{-10/7}$ energy decay rate and an 
$\ell(t)\sim t^{2/7}$ integral length scale growth rate that are consistent with 
assuming that an invariant exists that goes as $\cI\sim \overline{u}^2\ell^5$.
\item[$\circ$] Figure \ref{fig:r2d05EtINTG}b shows us that 
to understand the scaling for $\ell$ will require a better
understanding of the role of the $\eta k\sim0.01$ wavenumbers along with
their role in determining the circulation $\Gamma$. 
\item The persistence of the -5/3 spectra and 
the energy decay rate are consistent with those observed by a recent experiment 
following the decay of isotropic turbulence. With 
the differences in the growth rates of the respective integral scales $\ell$
possibly due to the effect of the largest scales, or $k\sim0.01$ wavenumbers.
\item[$\circ$] For times between $t=5.2$ and 30.4, a careful examination of how braids 
are shed by one ring, then wrap around the other, might shed light upon how
the Kolmogorov spectra are maintained over these times.
\item  That $\cH(t)/\sqrt{2E(t)Z(t)}$ is relatively constant in 
figure \ref{fig:KSRZHr2d05sqnu}b would support the existence of an invariant $\cI$ 
that could be related to the helicity $\cH(t)$, or something with its dimensions.
With one possible example coming from the streamlines shown in the inset of 
figure \ref{fig:3Dstruc}a, whose
associated circulations up to $t=10$ are $\Gamma\sim2$. 
\item[$\circ$] Additional measures for the circulation $\Gamma$ need to be considered.
\item[*] In addition to complementing the \cite{MatsuetalPNAS2026} experiment, 
these results should encourage further simulations using a collection 
of colliding vortex rings that would resemble the experiments more closely.
\ITN
\noindent {\small {\bf Acknowledgements} The author is grateful for discussions with 
P.A. Davidson and J.D. Gibbon. This work was partially supported by EPSRC 
small grant no. R.MAAA.3656.STB with computing resources provided by the 
Scientific Computing Research Technology Platform at the University of Warwick.}

\bibliographystyle{jfm}

\newpage

\end{document}